\documentclass[12pt]{iopart}
\usepackage{amssymb,cite,graphicx}

\def\k{{\bf k}}

\begin{document}

\title[Are there monojets in high-energy proton--nucleus collisions?]%
  {Are there monojets in high-energy\\ proton--nucleus collisions?}

\author{\underline{N Borghini}\dag\ and F Gelis\ddag}

\address{\dag\ Fakult\"at f\"ur Physik, Universit\"at Bielefeld, 
  Postfach 100131, D-33501 Bielefeld, Germany}

\address{\ddag\ Service de Physique Th\'eorique, CEA-Saclay, 
  F-91191 Gif-sur-Yvette cedex, France}

\begin{abstract}
We study high-energy proton--nucleus collisions within the Colour Glass 
Condensate framework, and compute the probabilities of having a definite number 
of scatterings in the nucleus with a momentum transfer larger than a given cut. 
Various properties of the distribution in the number of multiple scatterings are
investigated, and we conclude that events with monojets are very unlikely, 
except for extreme values of the saturation scale $Q_s$.
\end{abstract}



\vspace{-4.9mm}
\section{Introduction}
\label{s:intro}

High-energy collisions of hadrons probe partons with very small  $x$, 
whose occupation number becomes large, giving rise to the phenomenon of parton 
saturation~\cite{Mueller:1985wy}. 
The saturation is enhanced in a large nucleus, in which the nucleon wave 
functions overlap, and is expected to be revealed in high-energy collisions
involving such a heavy nucleus. 

Different frameworks implementing parton saturation have been developed, 
especially the McLerran--Venugopalan (MV) model~\cite{McLerran:1993ni,%
  McLerran:1993ka} and later the so-called ``Colour Glass Condensate'' 
(CGC)~\cite{Iancu:2003xm}.
A general feature in these approaches is the description of small-$x$ partons 
as classical Yang--Mills fields, radiated by the static large-$x$ colour 
sources. 
Computations of particle production in the collision of two saturated nuclei 
are then rather involved~\cite{Gelis:2007} and require numerical approaches. 
However, in interactions involving one small projectile, the colour sources 
describing the latter can be treated as weak: particle production can then be 
calculated by considering the relevant amplitude only at lowest order in this 
source, which allows one to obtain analytical expressions~\cite{%
  Jalilian-Marian:2005jf}. 

Hereafter, we shall focus on single-gluon production~\cite{Blaizot:2004wu}. 
While it is known that the CGC framework automatically includes all multiple 
scatterings that lead to the production of a gluon with a high transverse 
momentum $k_\perp$, we shall study in further detail the distribution in the 
number of these scatterings~\cite{Borghini:2006xh}, and more specifically, the 
scatterings with a momentum transfer larger than an arbitrary threshold 
$k_\perp^{\min}$.

\section{Models for the proton--nucleus interaction}

The number of gluons produced per unit of transverse momentum and per unit of 
rapidity in the collision between a proton and a saturated nucleus~\cite{%
  Blaizot:2004wu} involves the Fourier transform $C(\k_\perp)$ of a correlator 
of Wilson lines~\cite{Gelis:2001da}. 
When the distribution of colour sources inside the nucleus only has Gaussian 
correlations, as is the case in the two models we shall consider hereafter, this
correlator can be written in a form that has an interpretation in terms of 
independent multiple scatterings \`a la Glauber.
In particular, the contribution to the gluon yield of cases in which the 
outgoing gluon has undergone exactly $n$ scatterings can be isolated. 
One can thus show that the conditional probability that a gluon that comes out 
with a momentum $\k_\perp$ has undergone $n$ scatterings with momentum transfers 
larger than a threshold $k_\perp^{\min}$ (and an arbitrary number of softer 
scatterings) reads~\cite{Borghini:2006xh}
\begin{eqnarray}
\fl P_n(\k_\perp|k_\perp^{\min}) =
\frac{{\rm e}^{-\mu_0^2 \sigma_{\rm tot}}}{C(\k_\perp)}
\sum_{p=0}^{+\infty} \rho^{p+n} \int\limits_0^L\!{\rm d}z_1
\int\limits_{z_1}^L\!{\rm d}z_2\cdots\int\limits_{z_{p+n-1}}^L\!{\rm d}z_{n+p}
\int\limits_{\Lambda}^{k_\perp^{\min}}\!
\frac{{\rm d}^2{\k_1}_\perp}{(2\pi)^2}\cdots\frac{{\rm d}^2{\k_p}_\perp}{(2\pi)^2}
(2\pi)^2\nonumber\\
\times\!\int\limits_{k_\perp^{\min}}\!\!
 \frac{{\rm d}^2{\k_{p+1}}_\perp}{(2\pi)^2}\cdots
  \frac{{\rm d}^2{\k_{p+n}}_\perp}{(2\pi)^2}\,
\delta({\k_1}_\perp\!+\cdots+{\k_{p+n}}_\perp\!-\k_\perp)\,
\sigma({\k_1}_\perp)\cdots\sigma({\k_{p+n}}_\perp),\nonumber\\[-6mm]
\label{eq:Pn}
\end{eqnarray}
where $\rho$ is the number density of scattering centres in the nucleus, $L$ is 
the longitudinal size of the nucleus, $\mu_0^2\equiv \rho L$ is the density of
scattering centres per unit of transverse area, and $\sigma(\k_\perp)$ is the 
differential cross-section of a gluon with a scattering centre (the integral of 
which is $\sigma_{\rm tot}$).
We shall present results obtained with two versions of this cross-section: 
either that, proportional to $k_\perp^{-4}$, of the MV model --- in which the 
colour sources have a local Gaussian distribution ---, or the slightly more 
complicated cross-section, 
$\propto (Q_s^2/k_\perp^2)\ln[1\!+\!(Q_s^2/k_\perp^2)^\gamma]$ with 
$\gamma\approx 0.64$, of the non-local Gaussian effective theory (``asymptotic 
model'') that describes the gluonic content of a nucleus evolved to very small 
$x$ values~\cite{Iancu:2002aq}.
These two models also yield different relationships between the saturation scale
$Q_s$ and the scattering-centre density $\mu_0^2$ in equation~(\ref{eq:Pn}).

\section{Results}

\begin{figure}[t]
\includegraphics*[width=0.495\linewidth]{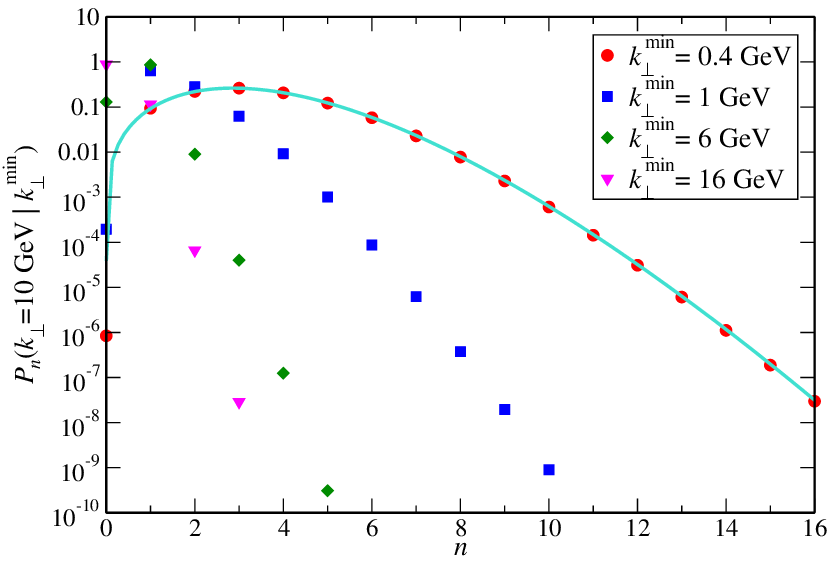}\hfill
\includegraphics*[width=0.495\linewidth]{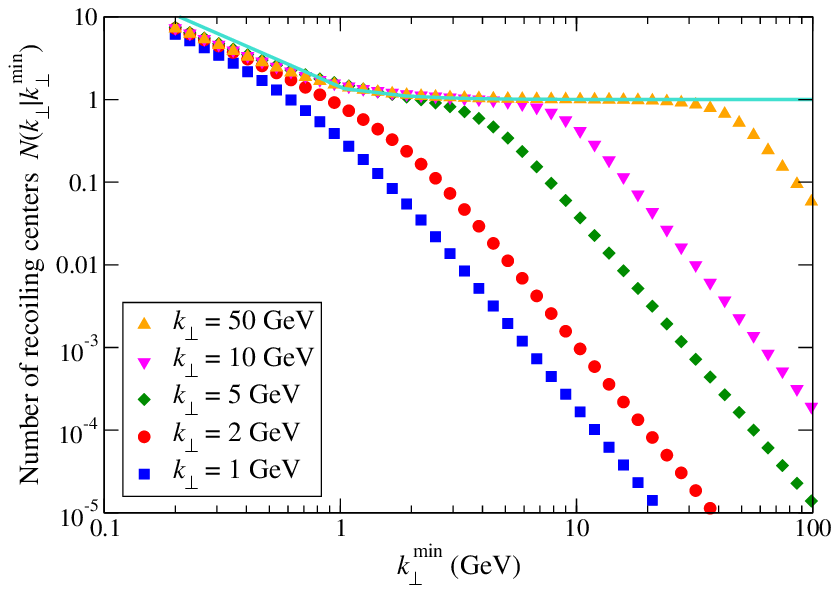}%
\caption{Left: distribution of the probabilities to have $n$ recoils when the 
  produced particle acquires $k_\perp\!=\!10$~GeV, for various values of the 
  threshold $k_\perp^{\min}$. 
  Right: number of recoils vs.\ the threshold $k_\perp^{\min}$, for various values
  of the acquired momentum $k_\perp$.
  Both panels assume the MV model with $Q_s^2=2\mbox{ GeV}^2$.}
\label{fig:fig1}
\end{figure}

Instead of computing directly the probability $P_n$, equation~(\ref{eq:Pn}), 
for each successive integer $n$, we rather calculate the generating function
\begin{equation}
F(z,\k_\perp|k_\perp^{\min}) \equiv \sum_{n=0}^{+\infty}P_n(k_\perp|k_\perp^{\min})\,z^n,
\label{eq:F}
\end{equation}
from which it is easy to extract the individual probabilities.
This function, which has a quite simple analytic expression (equation~(8) in 
reference~\cite{Borghini:2006xh}), also gives easy access to the average number 
of scatterings above the threshold $k_\perp^{\min}$, defined as
\begin{equation}
N(k_\perp|k_\perp^{\min})\equiv\sum_{n=1}^{+\infty} n P_n(k_\perp|k_\perp^{\min}),
\label{eq:N}
\end{equation}
which is obviously given by the value at $z=1$ of the first derivative of 
equation~(\ref{eq:F}).
The distribution of the probabilities $P_n(k_\perp|k_\perp^{\min})$ and the average 
number of recoils $N(k_\perp|k_\perp^{\min})$, computed within the MV model, are 
displayed in figure~\ref{fig:fig1}.
First, the width of the multiplicity distribution decreases with increasing 
$k_\perp^{\min}$, which is quite natural, since it is less and less likely to have
events in with a large number of recoils when $k_\perp^{\min}$ increases. 
One also sees that for all values of the threshold such that 
$Q_s\ll k_\perp^{\min}\lesssim k_\perp$, the most likely number of recoils is $n=1$,
while $n=0$ is most likely when $k_\perp<k_\perp^{\min}$.
This is paralleled by the fact that the number of recoils is very close to one
for any value of $k_\perp^{\min}$ such that $Q_s\ll k_\perp^{\min}\lesssim k_\perp$: 
when the gluon acquires a large momentum $k_\perp$, there is always one, and 
only one, hard recoil that provides most of this momentum.
On the other hand, $N(k_\perp|k_\perp^{\min})$ grows significantly at small 
$k_\perp^{\min}$, where it does not depend on $k_\perp$. 
This behaviour can be understood, since the generating function can be computed 
analytically when $k_\perp$ is much larger than the other two scales $Q_s$ and 
$k_\perp^{\min}$. 
One then finds that the probabilities $P_n$ follow a Poisson distribution 
shifted by one, corresponding to the compulsory hard scattering that provides 
most of the acquired momentum; the average value of this distribution is 
controlled by $Q_s$ and $k_\perp^{\min}$ only, independent of $k_\perp$.
This analytic prediction is represented by the solid line in the figures.

\begin{figure}[t]
\includegraphics*[width=0.495\linewidth]{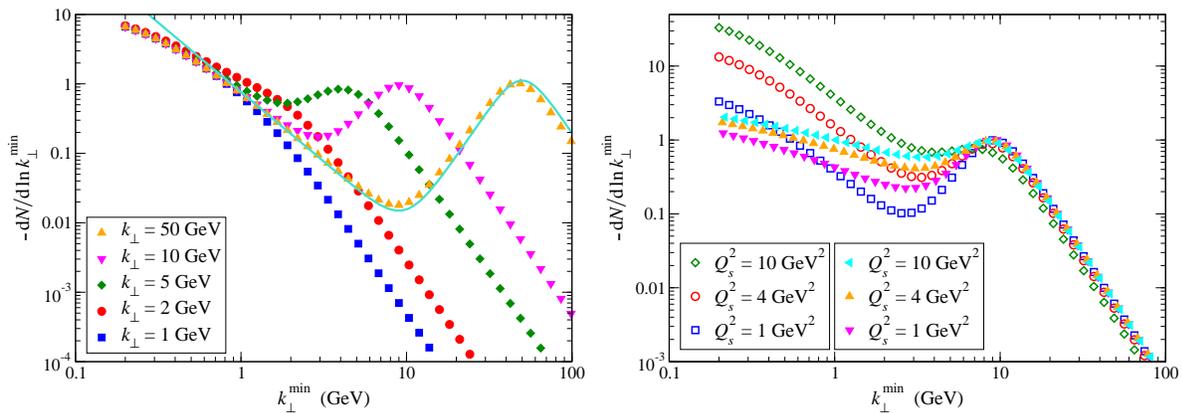}\hfill
\includegraphics*[width=0.495\linewidth]{fig2b}%
\caption{Distribution of the number of recoiling centres vs.\ the threshold 
  $k_\perp^{\min}$.
  Left: for various values of the acquired momentum $k_\perp$, within the MV 
  model with $Q_s^2=2\mbox{ GeV}^2$.
  Right: dependence on the saturation $Q_s$ within the MV (open symbols) and 
  asymptotic (full symbols) models, for a fixed value $k_\perp=10\mbox{ GeV}$.}
\label{fig:fig2}
\end{figure}

The interpretation in terms of a single hard scattering that provides most of 
the momentum is further supported by the momentum distribution of the recoils, 
left panel of figure~\ref{fig:fig2}. 
Indeed, this distribution consists of a universal semi-hard component made of 
recoils with momenta $\lesssim Q_s$, and a second component peaked around 
$k_\perp^{\min}=k_\perp$. 
In other words, the most likely configuration is that of a di-jet, the 
high-momentum gluon with $k_\perp$ being balanced by a second hard parton with 
similar momentum, rather than a monojet, which would correspond to having many 
soft partons on the away side. 

When the saturation scale $Q_s$ is varied (figure~\ref{fig:fig2}, right panel), 
the second, hard component around $k_\perp^{\min}=k_\perp$ is unchanged --- as it 
should, since it is controlled by hard physics, not by saturation. 
On the other hand, the semi-hard region is affected by changes in $Q_s$, with 
more and more scatterings as $Q_s$ increases. 
This growth in the number of soft recoils, which is quite spectacular within 
the MV model (open symbols in figure~\ref{fig:fig2}), is however significantly 
reduced in the asymptotic model (full symbols). 
This lesser sensitivity is believed to reflect the presence of leading-twist 
shadowing in the latter model: first, the shadowing somehow ``hides'' the 
scattering centres from the passing gluon. 
In addition, since the effect of piling up more and more colour sources in the 
nucleus is tamed by the shadowing, the dependence on $Q_s$ is weaker in the 
asymptotic model than in the shadowing-free MV model. 

Note also that the two-component structure of the momentum distribution of the 
recoils is less marked in the asymptotic model than in the MV model 
(figure~\ref{fig:fig2}, right panel). 
Especially, the dip between the two regions is filled up. 
This means that one should expect that the distribution of momenta in the 
``away-side jet''  will become flatter as one probes the nucleus at smaller and 
smaller $x$ values. 
The di-jet structure, however, still persists, unless the value of the 
saturation momentum $Q_s$ becomes very large, opening the possibility of having 
a hard jet balanced by a large number of softer particles on the other side.

\end{document}